\documentclass[reprint,aps,prd,superscriptaddress,showkeys,showpacs]{revtex4-1}
\usepackage{epsfig,amsmath,natbib}

\usepackage{aas_macros}
\usepackage{amssymb}
\usepackage{amsmath}
\usepackage{dsfont}
\usepackage{hyperref}
\usepackage{color}
\usepackage{pbox}
\usepackage{booktabs}
\usepackage[dvipsnames]{xcolor}

\hypersetup{
	colorlinks=false,
	citecolor=green
}

\newcommand{\bb}[1]{\mathbf{#1}}
\newcommand{\bbh}[1]{\mathbf{\hat{#1}}}
\newcommand{\h}[1]{\hat{#1}}

\newcommand{\ttt}[1]{\texttt{#1}}

\newcommand\lsim{\mathrel{\rlap{\lower4pt\hbox{\hskip1pt$\sim$}}
        \raise1pt\hbox{$<$}}}
\newcommand\gsim{\mathrel{\rlap{\lower4pt\hbox{\hskip1pt$\sim$}}
        \raise1pt\hbox{$>$}}}

\newcommand\pt{\pmb{\theta}}

\begin{document}

\title{On the validity of the Born approximation for beyond-Gaussian weak lensing observables}

\author{Andrea Petri}
\email{apetri@phys.columbia.edu}
\affiliation{Department of Physics, Columbia University, New York, NY 10027, USA}
\affiliation{Physics Department, Brookhaven National Laboratory, Upton, NY 11973, USA}

\author{Zolt\'an Haiman}
\affiliation{Department of Astronomy, Columbia University, New York, NY 10027, USA}
\affiliation{Department of Physics, New York University, New York, NY 10003, USA}

\author{Morgan May}
\affiliation{Department of Physics, Columbia University, New York, NY 10027, USA}
\affiliation{Physics Department, Brookhaven National Laboratory, Upton, NY 11973, USA}

\date{\today}

\label{firstpage}

\begin{abstract}
Accurate forward modeling of weak lensing (WL) observables from cosmological parameters is necessary for upcoming galaxy surveys. Because WL probes structures in the non--linear regime, analytical forward modeling is very challenging, if not impossible. Numerical simulations of WL features rely on ray--tracing through the outputs of $N$--body simulations, which requires knowledge of the gravitational potential and accurate solvers for light ray trajectories. A less accurate procedure, based on the Born approximation, only requires knowledge of the density field, and can be implemented more efficiently and at a lower computational cost. In this work, we use simulations to show that deviations of the Born--approximated convergence power spectrum, skewness and kurtosis from their fully ray--traced counterparts are consistent with the smallest non--trivial $O(\Phi^3)$ post--Born corrections (so-called geodesic and lens-lens terms). Our results imply a cancellation among the larger $O(\Phi^4)$ (and higher order) terms, consistent with previous analytic work. We also find that cosmological parameter bias induced by the Born approximated power spectrum is negligible even for an LSST--like survey, once galaxy shape noise is considered. When considering higher order statistics such as the $\kappa$ skewness and kurtosis, however, we find significant bias of up to 2.5$\sigma$. Using the \ttt{LensTools} software suite, we show that the Born approximation saves a factor of 4 in computing time with respect to the full ray--tracing in reconstructing the convergence.       
\end{abstract}


\maketitle


\section{Introduction}
Weak gravitational lensing (WL) is a promising observational technique to probe the standard $w$CDM model of the universe, and can help in constraining the Dark Energy equation of state parameters \citep{wlreview}. Accurate forward modeling of WL observables is crucial for parameter estimation purposes. This however is a challenging problem due to the non--linear nature of WL image fields. Numerical efforts for feature forward modeling are based on ray--tracing simulations that make use of the multi--lens--plane algorithm \citep{RayTracingJain,RayTracingHartlap,RayTracingPN}, or the direct approach \citep{RayTracingBecker,RayTracingBode,RayTracingBarreira}, to predict cosmic shear from outputs of $N$--body simulations. Approximate techniques based on the Born approximation allow for a computationally faster, but potentially inaccurate, forward modeling which requires only knowledge of the matter density contrast integrated along the observer unperturbed line of sight. This has been done both in the flat sky \citep{RayTracingHartlap} and full sky \citep{Fosalba1,Fosalba2} limits. A variety of groups have studied the validity of the Born approximation for forward modeling of convergence power spectra \citep{HirataKrause} and bi--spectra \citep{WLBispectrumDodelson}, spectra of WL cosmic flexions \citep{BornFlexion} and, in more recent work, CMB lensing bi--spectra \citep{CMBPrattenLewis}. In this work, we study the validity of the Born approximation for forward modeling a subset of low--order moments of the real--space convergence field. 

This paper is organized as follows.  In \S\ref{sec:sims}, we give a brief review of the ray--tracing formalism, along with the Born approximation and its lowest--order corrections. In \S~\ref{sec:params}, we then outline how cosmological parameter constraints can be derived from WL features. We present our main results in \S~\ref{sec:results}, which we discuss further in \S~\ref{sec:discuss}.  Finally, in \S~\ref{sec:conclude}, we summarize our conclusions, and mention possible future extensions of this study.       


\section{Simulations}
\label{sec:sims}
In this section, we give an overview of the formalism behind WL simulations. WL observables are related to the matter density contrast $\delta$ (or potential $\Phi$). Light rays crossing density inhomogeneities experience deflections, which cause observed galaxy shapes to be distorted. Modeling of image distortions can be done by computing light ray geodesics in the density field $\delta$. To simulate the matter density field we make use of the public code \ttt{Gadget2} \citep{Gadget2} and we run $N$--body simulations with a box size of $L_b=260\,{\rm Mpc}/h$ and $N_p=512^3$ particles, which corresponds to a mass resolution of $M_p\approx 10^{10}M_\odot$ per particle. We then perform a grid--based density estimation based on the position of the particles at different times. From the three dimensional density contrast $\delta$, the two dimensional lensing potential $\Phi$ can be inferred by solving the Poisson equation

\begin{equation}
\label{sim:poisson}
\nabla_\perp^2\Phi(\bb{x}_\perp,\chi) = \frac{3H_0^2\Omega_m}{2c^2a(\chi)} \delta(\bb{x}_\perp,\chi).
\end{equation} 
Where we denote the transverse comoving coordinates as $\bb{x}_\perp$, the longitudinal comoving distance as $\chi$, the universe scale factor as $a$, and the Hubble parameter and the matter density parameter as  $H_0$ and $\Omega_m$, respectively. The time dependence of $\Phi,\delta$ has been absorbed into $\chi$ with the use of the distance--redshift relation. Under the flat--sky approximation, the transverse coordinates $\bb{x}_\perp$ can be related as the angles as seen from an Earth based observer $\pmb{\beta}=\bb{x}_\perp/\chi$. In the following paragraph we give a summary of the ray--tracing basics.   

\subsection{Ray tracing}
Light ray trajectories correspond to null geodesics in the space--time metric induced by the density fluctuations $\delta(\bb{x}_\perp,\chi)$. It can be shown that with the spacetime parametrization adopted in this section, the geodesic equation takes the form (see \citep{BornFlexion})

\begin{equation}
\label{sim:geodiff}
\frac{d^2 \bb{x}_\perp(\chi)}{d\chi^2} = -2\nabla_\perp \Phi(\bb{x}_\perp,\chi)
\end{equation}
Equation (\ref{sim:geodiff}) can be directly integrated to express its solution $\bb{x}_\perp(\chi)$ in explicit form (see \citep{DodelsonWL}) 

\begin{equation}
\label{sim:geosol}
\bb{x}_\perp(\chi_s) = \bb{x}_\perp(0) -2\int_0^{\chi_s} d\chi (\chi_s-\chi) \nabla_\perp \Phi(\bb{x}_\perp(\chi),\chi)
\end{equation}
Where $\chi_s$ is the source galaxy comoving distance. WL observables are related to the differential deflection that nearby light rays experience when traveling from the observer to the source. Indicating with $\pt$ the starting angular position of a light ray at $\chi=0$ and with $\pmb{\beta}_s$ the angular position of the light ray at $\chi=\chi_s$, we are interested in the Jacobian matrix $\bb{A}_s(\pt)=\partial\pmb{\beta}_s(\pt)/\partial \pt$, which can be parametrized as 

\begin{equation}
\label{sim:lensjac}
\bb{A}_s(\pt) = 
\begin{pmatrix}
1-\kappa(\pt)-\gamma_1(\pt) & -\gamma_2(\pt) + \omega(\pt) \\
-\gamma_2(\pt) - \omega(\pt) & 1-\kappa(\pt)+\gamma_1(\pt)
\end{pmatrix}
\end{equation}
Where $\kappa$, and $\pmb{\gamma}$ refer respectively to the WL
convergence and shear, and $\omega$ to the rotation angle (which is
$O(\Phi^2)$ and is often excluded from the parametrization in eq.~(\ref{sim:lensjac})).  The implicit solution
(\ref{sim:geosol}) to the geodesic equation (\ref{sim:geodiff}) can be
translated in an integral equation for the lensing Jacobian

\begin{widetext}
\begin{equation}
\label{sim:jacsol}
A_{ij}(\pt,\chi_s) = \delta^K_{ij}-2\int_0^{\chi_s} d\chi\chi W(\chi,\chi_s)\Phi_{ik}(\bb{x}_\perp(\chi,\pt),\chi)A_{kj}(\pt,\chi).
\end{equation} 
\end{widetext}
We denote the partial derivatives of the lensing potential $\Phi$ with respect to the transverse coordinates $x,y$ as subscripts. We also defined the single--redshift source lensing kernel $W(\chi,\chi_s)\equiv 1-\chi/\chi_s$ and indicated the Kronecker delta symbol as $\delta^K$. The WL convergence $\kappa(\pt,\chi_s)$ can be calculated from the trace of the Jacobian
\begin{equation}
\label{sim:trkappa}
\kappa(\pt,\chi_s) = 1-\frac{1}{2}{\rm tr}\left[\bb{A}(\pt,\chi_s)\right]
\end{equation}
The implicit form of equation (\ref{sim:jacsol}) suggests a straightforward way to solve the geodesic equation numerically because $A(\pt,\chi_s)$ can be calculated once $A(\pt,\chi)$ for $\chi<\chi_s$ is known. The multi--lens--plane algorithm \citep{RayTracingJain,RayTracingHartlap} is a popular method to compute the integral in (\ref{sim:jacsol}) in discrete steps (lenses) by keeping track of the intermediate values $A(\pt,\chi)$ using dynamic programming. To carry out the geodesic calculations we make use of the \ttt{LensTools} software package \citep{LensTools-paper}, which provides a {\sc python} implementation of the multi--lens--plane algorithm. The exact solution of (\ref{sim:geodiff}) based on (\ref{sim:jacsol}) is computationally expensive because it requires knowledge of the lensing potential $\Phi$. In the next paragraph we review the details of a computationally faster but approximate approach based on the Born approximation.     

\subsection{Born approximation}
The integral in (\ref{sim:jacsol}) can be approximated using a series expansion in powers of $\Phi$, which can yield acceptable results in the limit in which $\Phi$ is small. To compute the lowest--order approximation to eq.~(\ref{sim:jacsol}) one notes that the lensing potential appears on the right hand side and hence we can replace the Jacobian on the right hand side with its zeroth order expression, i.e. the identity matrix. We can also replace the real ray trajectory $\bb{x}_\perp(\chi,\pt)$ with the unperturbed one, i.e. $\chi\pt$. This yields an expression for $\kappa$ at first order in $\Phi$

\begin{equation}
\label{sim:born}
\kappa_{\rm born}(\pt,\chi_s) = \frac{3H_0^2\Omega_m}{2c^2}\int_0^{\chi_s} d\chi\frac{\chi}{a(\chi)} W(\chi,\chi_s)\delta(\chi\pt,\chi)
\end{equation}
The Born--approximated convergence in eq.~(\ref{sim:born}) has a simple interpretation: at lowest order in the lensing potential, $\kappa$ is the integrated matter density on the unperturbed line of sight, weighted by the lensing kernel $W$. Contrary to the exact ray--tracing approach, the Born approximation does not require knowledge of the solution to the Poisson equation (\ref{sim:poisson}) and is hence computationally faster. In the next paragraph we examine the corrections to the Born approximation that come at second order in the potential $\Phi$.  

\subsection{Post--Born corrections}

Equation (\ref{sim:jacsol}) allows us to express the convergence $\kappa$ at arbitrary high orders in $\Phi$ powers. The linear--order expression corresponds to the Born approximation in equation (\ref{sim:born}). At second order in $\Phi$ we can write, following \citep{WLBispectrumDodelson}

\begin{equation}
\label{sim:pb2}
\kappa = \kappa_{\rm born} + \kappa_{\rm ll} + \kappa_{\rm geo} + O(\Phi^3)
\end{equation}
The second order O($\Phi^2$) corrections can be expressed as double integrals along the unperturbed line of sight

\begin{figure*}
\begin{center}
\includegraphics[scale=0.4]{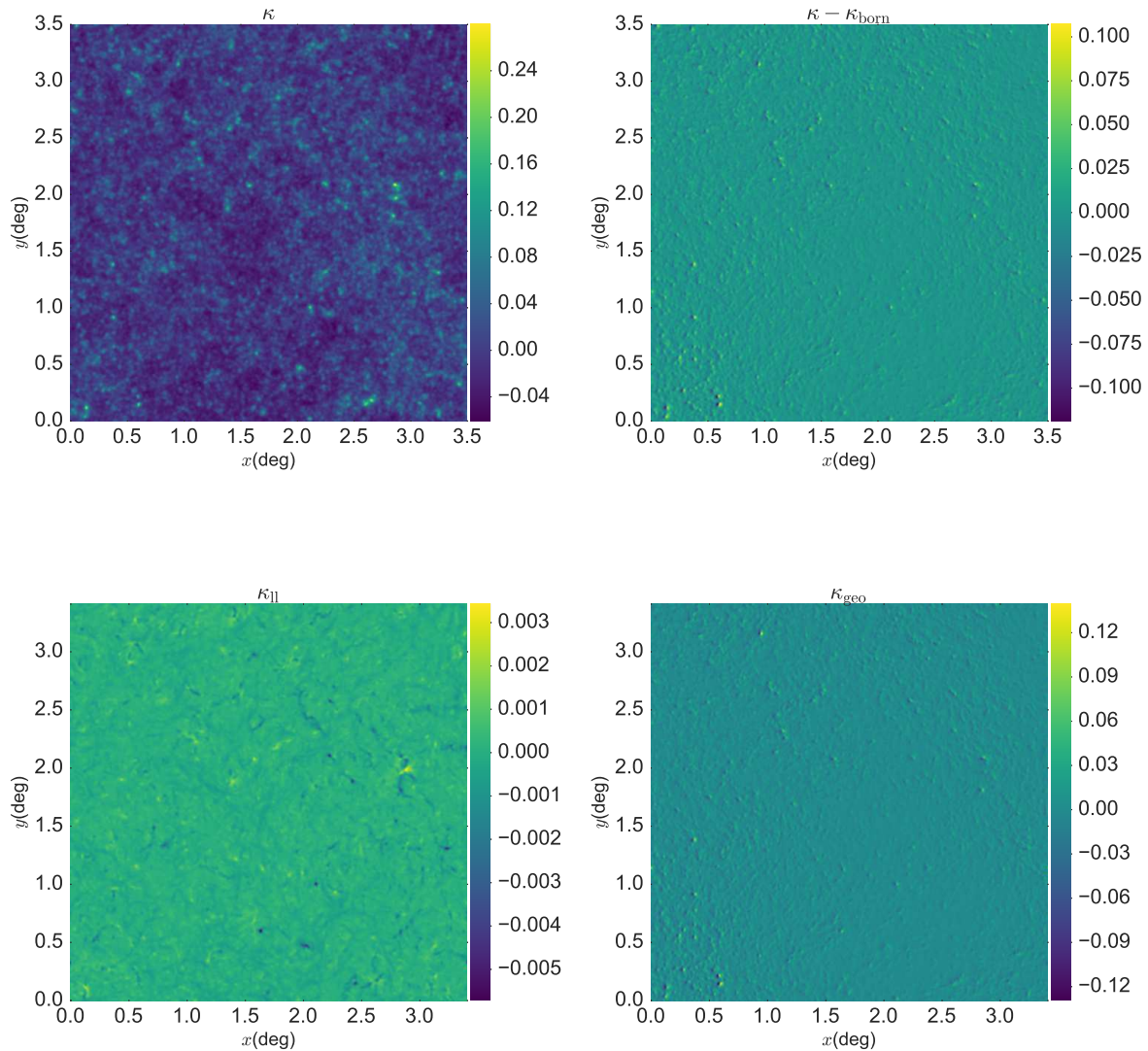}
\end{center}
\caption{Sample convergence outputs for one realization of a $(3.5\,{\rm deg})^2$ field, with source galaxies at redshift $z_s=2$. The figure shows the convergence profile (top left), along with the Born approximation residuals (top right), the lens-lens post--Born contribution (bottom left) and the geodesic contribution (bottom right). The images have been smoothed with a Gaussian kernel of size $\theta_G=0.5\,{\rm arcmin}$.  Note the different scales in the bottom panels: the geodesic term (right), is over an order of magnitude larger than the lens--lens term (left), and is comparable in level and detailed structure to the full residual (top right). The ``22'' and ``33'' post--Born corrections to the $\kappa$ power spectrum, which are widely mentioned in the literature \citep{HirataKrause} are not directly observable from these $\kappa$ images.} 
\label{fig:csample}
\end{figure*}

\begin{widetext}

\begin{equation}
\label{sim:ll}
\kappa_{\rm ll}(\pt,\chi_s) = -2\int_0^{\chi_s}d\chi\int_0^\chi d\chi' \chi\chi' W(\chi,\chi_s)W(\chi',\chi)\Phi_{ij}(\chi\pt,\chi)\Phi_{ij}(\chi'\pt,\chi'),
\end{equation}

\begin{equation}
\label{sim:gp}
\kappa_{\rm geo}(\pt,\chi_s) = -\frac{3H_0^2\Omega_m}{c^2}\int_0^{\chi_s}d\chi\int_0^\chi d\chi' \frac{\chi\chi'}{a(\chi')} W(\chi,\chi_s)W(\chi',\chi)\nabla_\perp\Phi(\chi\pt,\chi)\cdot\nabla_\perp\delta(\chi'\pt,\chi').
\end{equation}

\end{widetext}
These second--order expressions have a simple physical interpretation. Equation (\ref{sim:ll}) encodes the post--Born correction to the convergence due to non--local quadratic lens--lens couplings, which add to the density line of sight integral in eq.~(\ref{sim:born}). Eq.~(\ref{sim:gp}), on the other hand, encodes the integral of the density contrast along the real light--ray trajectory, at lowest order in the geodesic deflections. The line of sight integrals eqs.~(\ref{sim:born}), (\ref{sim:ll}) and (\ref{sim:gp}) can be computed efficiently with runtime that scales linearly with the number of lenses using the functionality of the \ttt{LensTools} suite \citep{LensTools-paper}. A sample of the simulation outputs is shown in Figure \ref{fig:csample}. Benchmarks for the runtime of the ray--tracing and line--of--sight integrals operations, performed with \ttt{LensTools}, are provided in Table \ref{tab:benchmarks}. In the next section, we summarize the procedure to fit forward models of the WL features to observational data to infer cosmological parameters.

\begin{table*}
\begin{center}

\begin{tabular}{c|c|c|c}

\textbf{Integration type} & \textbf{Runtime (One field of view)} & \textbf{Memory usage} & \textbf{CPU time (1000 fields of view)} \\ \hline \hline
Born & 36.0\,s & 0.86\,GB & 10\,hours  \\
Full ray--tracing & 124.8\,s & 1.65\,GB & 35\,hours  \\
Born + $O(\Phi^2)$ & 156.7\,s & 1.52\,GB & 44\,hours \\ \hline

\end{tabular}

\end{center}

\caption{CPU time and memory usage benchmarks for the convergence reconstruction operations. The test case we refer to in this table matches the specifications of the simulations used in this work. We we divide the line of sight in $N_l=42$ uniformly spaced lenses between the observer and the sources at $z_s=2$, each with a resolution of $4096^2$ pixels. The $\kappa$ field is resolved with $2048^2$ light rays. We show both the runtime for producing a single field of view and the CPU hours needed to perform the reconstruction 1000 times, to mock an LSST--like galaxy survey. Run times do not include the Poisson solution calculation, as this can be recycled to produce multiple field of view realizations. The Poisson solution run time is negligible in the account of the total CPU time needed for the production of $N_r\gg N_l$ field of view realizations.}
\label{tab:benchmarks}

\end{table*}

\section{Parameter inference}
\label{sec:params}
Parameter estimation from WL observations involves measurement of an image feature $\bbh{d}_{\rm obs}$ from a reconstructed convergence field $\h{\kappa}(\pt)$. The measured image feature is matched against a forward model $\bb{d}(\bb{p})$ to get an estimate of the cosmological parameters $\bbh{p}$. In the limit in which the feature likelihood is a multivariate Gaussian with $\bb{p}$--independent covariance $\bb{C}^{\bb{d}\bb{d}}$ and the forward model is linear in the parameters 

\begin{equation}
\label{par:linapprox}
\bb{d}(\bb{p}) = \bb{d}_0 + \bb{M}(\bb{p}-\bb{p}_0)
\end{equation}
the maximum likelihood parameter estimator is given by (see \citep{DodelsonSchneider13,PetriVariance} for example)
\begin{equation}
\label{par:fisherest}
\bbh{p} = \bb{p}_0 + (\bb{M}^T\bb{\Psi M})^{-1} \bb{M}^T\bb{\Psi} (\bbh{d}_{\rm obs}-\bb{d}_0)
\end{equation}
where $\bb{\Psi}=(\bb{C}^{\bb{d}\bb{d}})^{-1}$. The linearity assumption (eq.~\ref{par:linapprox}) is justified when the scatter of the measure feature $\bbh{d}_{\rm obs}$ around the expansion point $\bb{d}_0$ is expected to be small. This is true for upcoming large area surveys such as LSST \citep{LSST}, WFIRST \citep{WFIRST} and Euclid \citep{Euclid}. In this work we use $N$--body simulations coupled with the \ttt{LensTools} routines to simulate multiple realizations (using a sampling procedure analogous to the one outlined in \citep{PetriVariance}) of a convergence field $\h{\kappa}(\pt)$ for different combinations of the cosmological parameter triplet $(\Omega_m,w_0,\sigma_8)$. We choose the values $\bbh{p}_0=(0.26,-1,0.8)$ as our fiducial cosmological parameters and vary one parameter at a time as in Table \ref{tab:cosmopar} to measure the feature derivatives $\bb{M}$. 

To study the effect of the Born approximation on parameter constraints, we measure the fiducial features $\bb{d}_0$, the derivatives $\bb{M}$ and the feature covariance $\bb{C}^{\bb{d}\bb{d}}$ from $\kappa$ mocks constructed with the Born approximation (eq.~\ref{sim:born}). We then evaluate the induced parameter bias by applying the parameter estimator (eq.~\ref{par:fisherest}) to a mock observation constructed with full ray--tracing as in equations (\ref{sim:jacsol}) and (\ref{sim:trkappa}). We perform the ray--tracing and the line--of--sight integrals for sources at the  fixed redshift $z_s=2$, uniformly distributed on a square grid in a $(3.5\,{\rm deg})^2$ field of view, with a resolution of $2048^2$ pixels. To mimic an LSST--like mock observation, we use the 8192 realizations in our fiducial $\kappa$ ensemble to bootstrap the mean of 1000 fiducial feature measurements, which is equivalent approximately to a $10000\,{\rm deg}^2$ sky coverage. We use this bootstrapped mean as $\bbh{d}_{\rm obs}$.   

\begin{table}
\begin{center}

\begin{tabular}{c|c|c|c}

$\Omega_m$ & $w_0$ & $\sigma_8$ & $\kappa$ realizations \\ \hline \hline
\multicolumn{4}{c}{\textbf{Fiducial}} \\ \hline
0.26 & $-1$ & 0.8 & 8192 \\ \hline

\multicolumn{4}{c}{\textbf{Variations}} \\ \hline
0.29 & $-1$ & 0.8 & 1024 \\
0.26 & $-0.8$ & 0.8 & 1024 \\
0.26 & $-1$ & 0.9 & 1024 \\ \hline

\end{tabular}

\end{center}

\caption{Cosmological parameters in our simulation suite}
\label{tab:cosmopar}

\end{table}

We add galaxy shape noise to our $\kappa$ mocks in the form of a pixel uncorrelated white Gaussian noise \citep{SongKnox} with a root mean square of
\begin{equation}
\label{par:shaperms}
\sigma_{\rm shape} = \frac{0.15+0.035z_s}{\sqrt{N_g}}
\end{equation}
where $N_g$ is the number of galaxies per pixel. 

Our image feature choice includes the $\kappa$ power spectrum $P^{\kappa\kappa}(\ell)$, defined as 
\begin{equation}
\label{par:powerdef}
\langle\tilde{\kappa}(\pmb{\ell})\tilde{\kappa}(\pmb{\ell}')\rangle = (2\pi)^2\delta^D(\pmb{\ell}+\pmb{\ell}')P^{\kappa\kappa}(\ell),
\end{equation}
where the expectation value is taken over $\pmb{\ell}$ modes with the same magnitude $\ell=\vert\pmb{\ell}\vert$ and $\delta^D$ is the Dirac delta function. Because the $\kappa$ field is non--Gaussian, higher order statistics such as higher real space $\kappa$ moments have been shown to contain complementary information in addition to the one already supplied by the power spectrum \citep{MinkPetri,CFHTMink,NG-Jain1,NG-Jain2}. We consider the following sets of $\kappa$ moments $\pmb{\mu}_i$:

\begin{equation}
\label{par:moments}
\begin{matrix}
\pmb{\mu} = (\pmb{\mu}_2,\pmb{\mu}_3,\pmb{\mu}_4) \\ \\
\pmb{\mu}_2 = \left(\langle\kappa^2\rangle,\langle\vert\nabla\kappa\vert^2\rangle\right) \\ \\
\pmb{\mu}_3 = \left(\langle\kappa^3\rangle,\langle\kappa^2\nabla^2\kappa\rangle,\langle\vert\nabla\kappa\vert^2\nabla^2\kappa\rangle\right) \\ \\
\pmb{\mu}_4 = \left(\langle\kappa^4\rangle_c,\langle\kappa^3\nabla^2\kappa\rangle_c,\langle\kappa\vert\nabla\kappa\vert^2\nabla^2\kappa\rangle_c,\langle\vert\nabla\kappa\vert^4\rangle_c\right).\\
\end{matrix}
\end{equation} 
Here the expectation values are taken with respect to pixels and the subscript $c$ denotes the connected components of the quartic moments $\pmb{\mu}_4$. In the next section we outline our main results.  


\section{Results}
\label{sec:results}
In this section, we present the main results of this work, regarding accuracy of forward modeling the WL features, and the cosmological parameter bias induced by the Born approximation.

\begin{figure*}
\begin{center}
\includegraphics[scale=0.45]{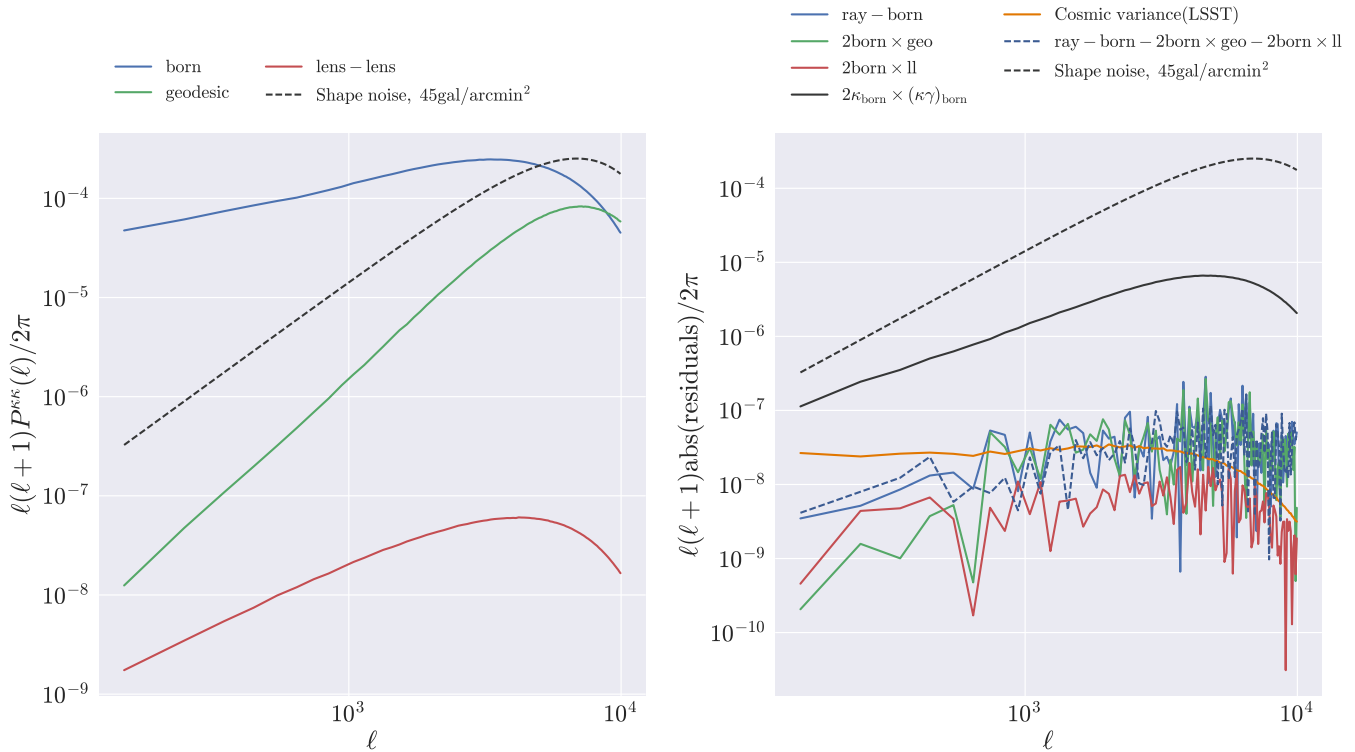}
\end{center}
\caption{Contributions at different $\Phi$ orders to the $\kappa$ power spectrum: the left panel shows the auto power spectra $P_{\rm born,born}$ ($O(\Phi^2)$) and $P_{\rm ll},P_{\rm geo}$ ($O(\Phi^4)$). The right panel shows the residuals between the power spectrum of the full ray--traced $\kappa$ field and the one obtained with the Born approximation. For reference, we show the shape noise contributions (black dashed) and the first non--trivial reduced shear corrections to $P^{\kappa\kappa}$ (black solid), which can be both added to the forward models after the line of sight integration. We also show as an orange line the level of cosmic variance for the power spectrum expected in a LSST--like survey. Gaussian convolution effects with a window of size $\theta_G=0.5\,{\rm arcmin}$ are included. The residuals are compared to the two $O(\Phi^3)$ terms $2P_{\rm born,ll},2P_{\rm born,geo}$. The quantities shown are the ensemble averages over 8192 realizations of $\kappa$ in the fiducial cosmology $\bb{p}_0$.}
\label{fig:psResiduals}
\end{figure*}

\begin{figure*}
\begin{center}
\includegraphics[scale=0.46]{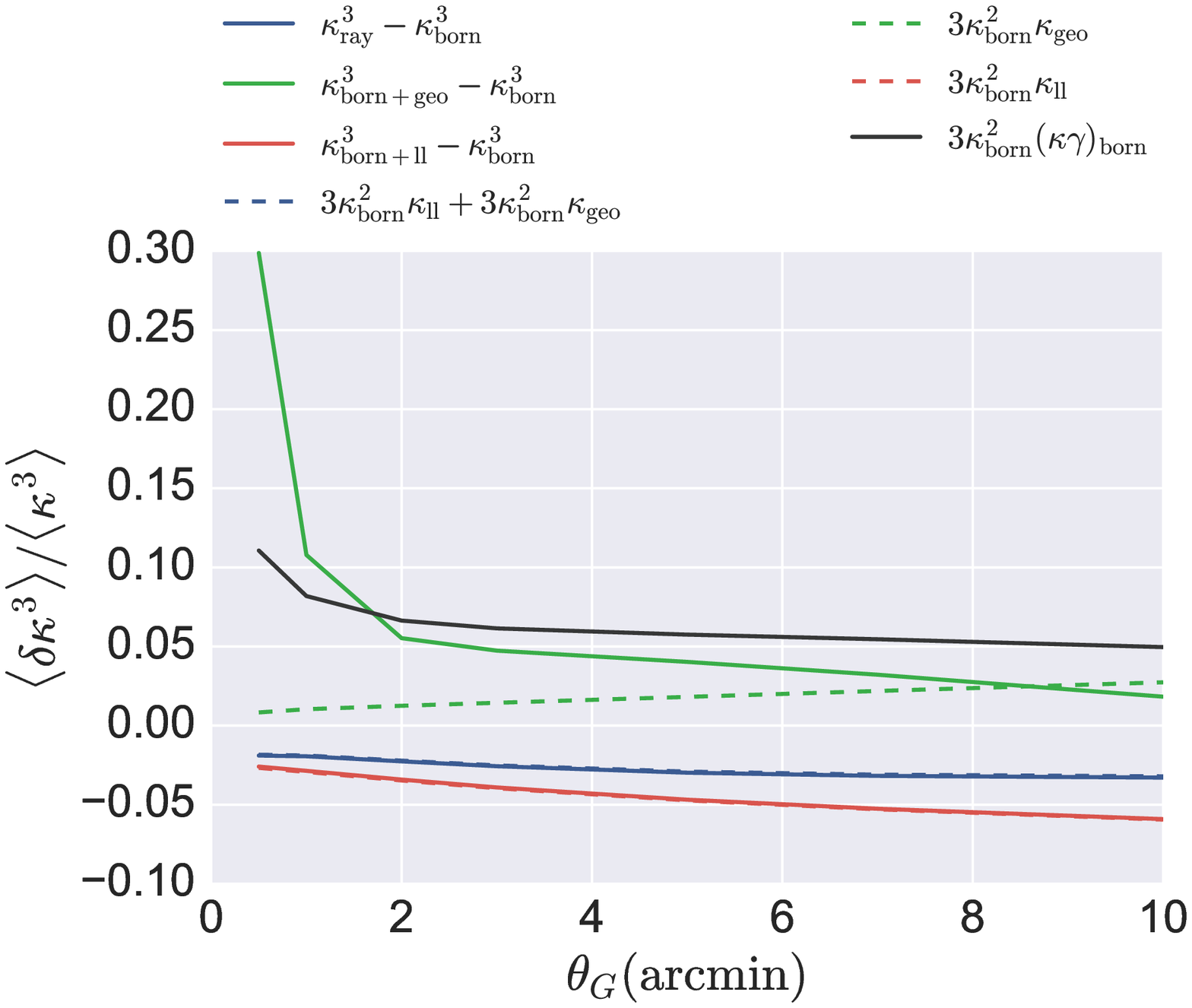}\includegraphics[scale=0.46]{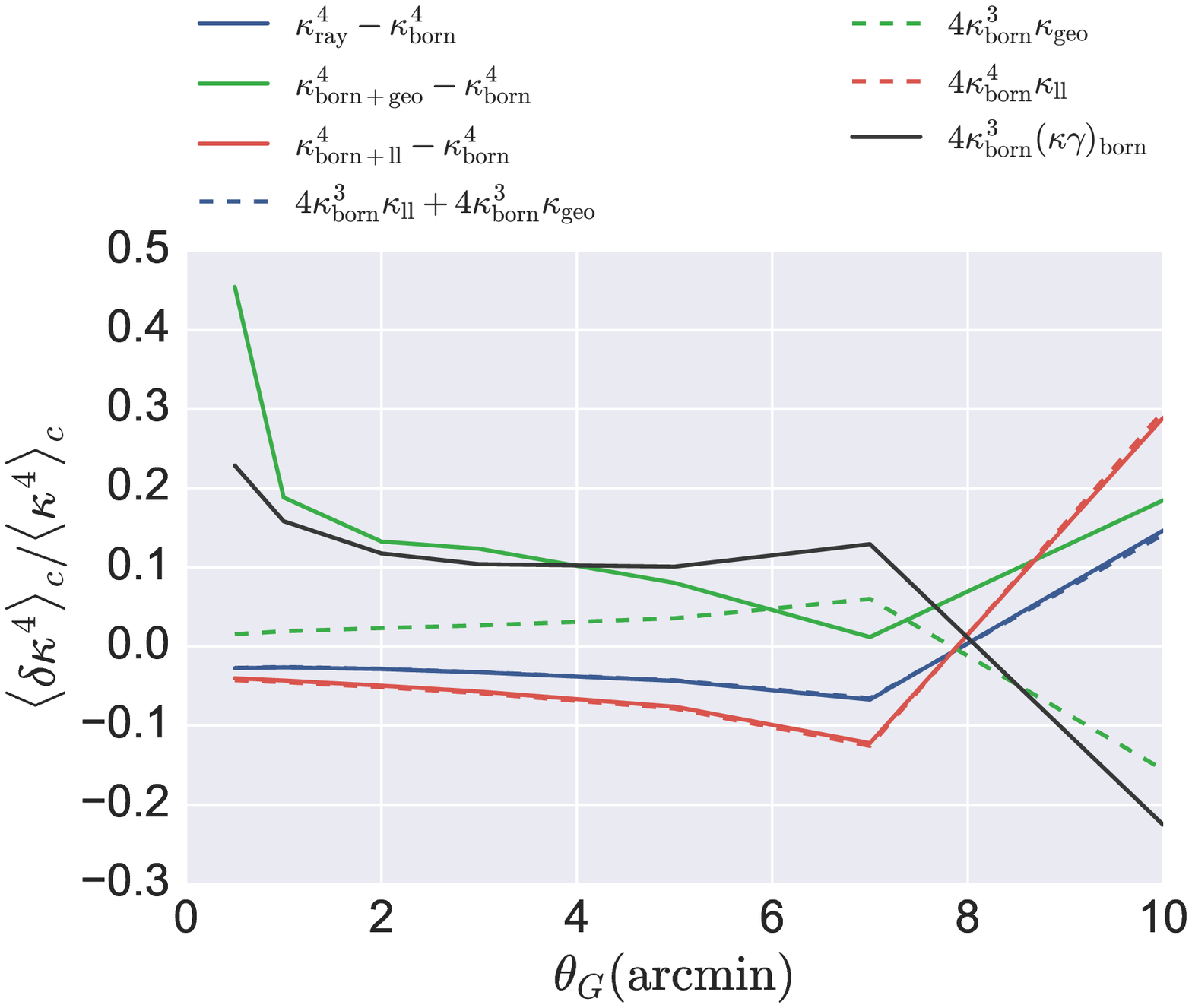}
\end{center}
\caption{Dominant post--Born corrections to the $\kappa$ skewness and kurtosis due to lens--lens coupling and geodesic perturbation, as a function of the scale $\theta_G$ of the Gaussian smoothing window. We show the contributions of the lens--lens couplings (red) and the geodesic perturbation (green). We show both the cases in which the residuals $\delta\kappa^3,\delta\kappa^4$ are computed adding equations (\ref{sim:ll},\ref{sim:gp}) to the Born field (solid lines) and in which the residuals are computed from the dominant $\Phi$ orders only as in equations (\ref{res:skewcross},\ref{res:kurtcross}) (dashed lines). For reference we show the first non--trivial corrections due to reduced shear (black lines), which can be calculated from the Born images. The quantities shown are the ensemble averages over 8192 realizations of $\kappa$ in the fiducial cosmology $\bb{p}_0$.}
\label{fig:skResiduals}
\end{figure*} 

\begin{figure*}
\begin{center}
\includegraphics[scale=0.3]{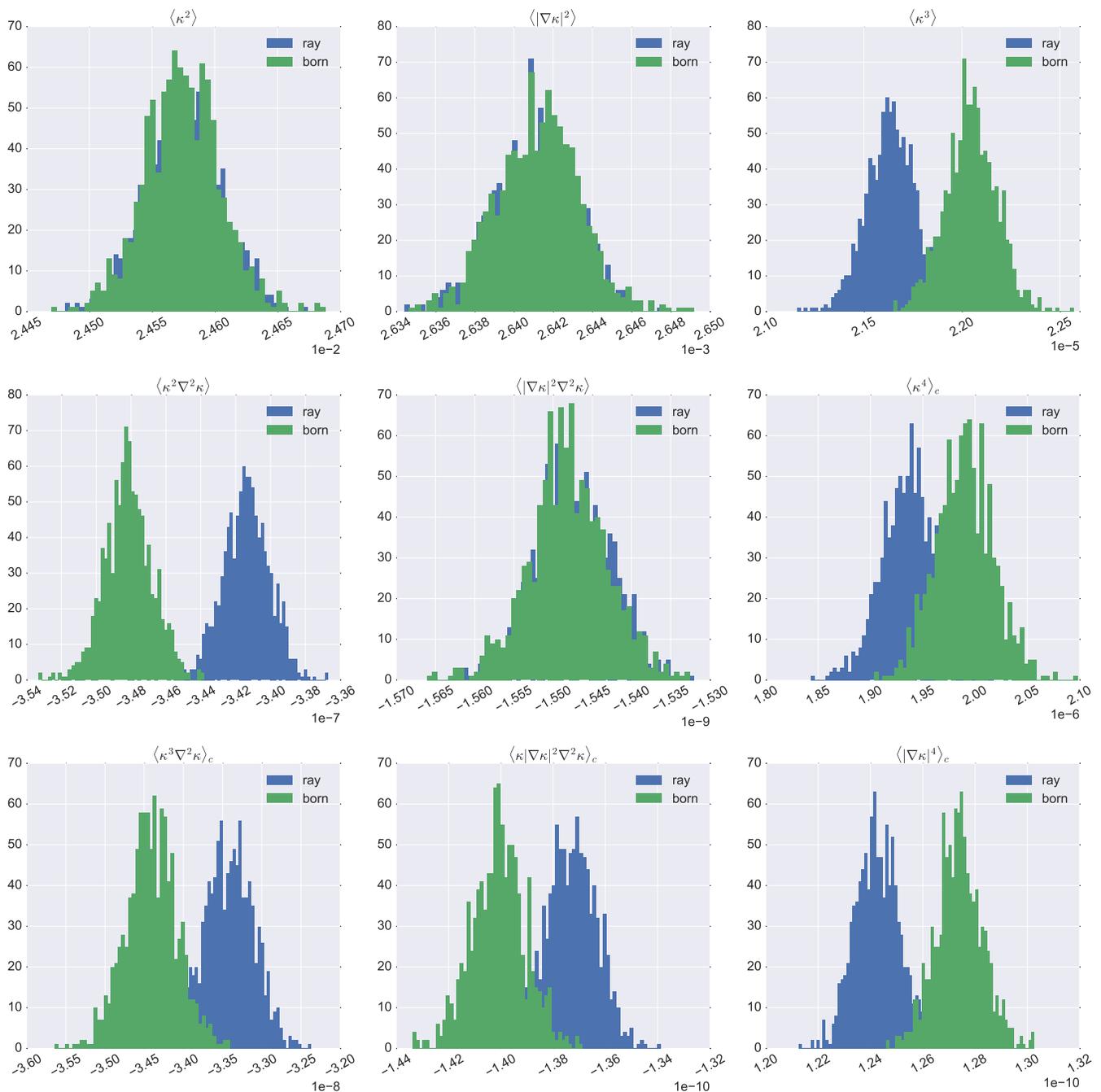}
\end{center}
\caption{Probability distribution function (PDF) of the 9 $\kappa$ moments defined in eq.~(\ref{par:moments}), over 8192 realizations of the fiducial cosmological model $\bb{p}_0$. The top/middle/bottom rows show examples of 2nd, 3rd, and 4th-order moments. We show both the PDFs obtained with full ray--tracing (blue) and the Born approximation (green). No shape noise has been added. The $\kappa$ maps were smoothed with a Gaussian kernel of size $\theta_G=0.5\,{\rm arcmin}$ before measuring the moments.}
\label{fig:pdfMoments}
\end{figure*} 

\begin{figure*}
\begin{center}
\includegraphics[scale=0.3]{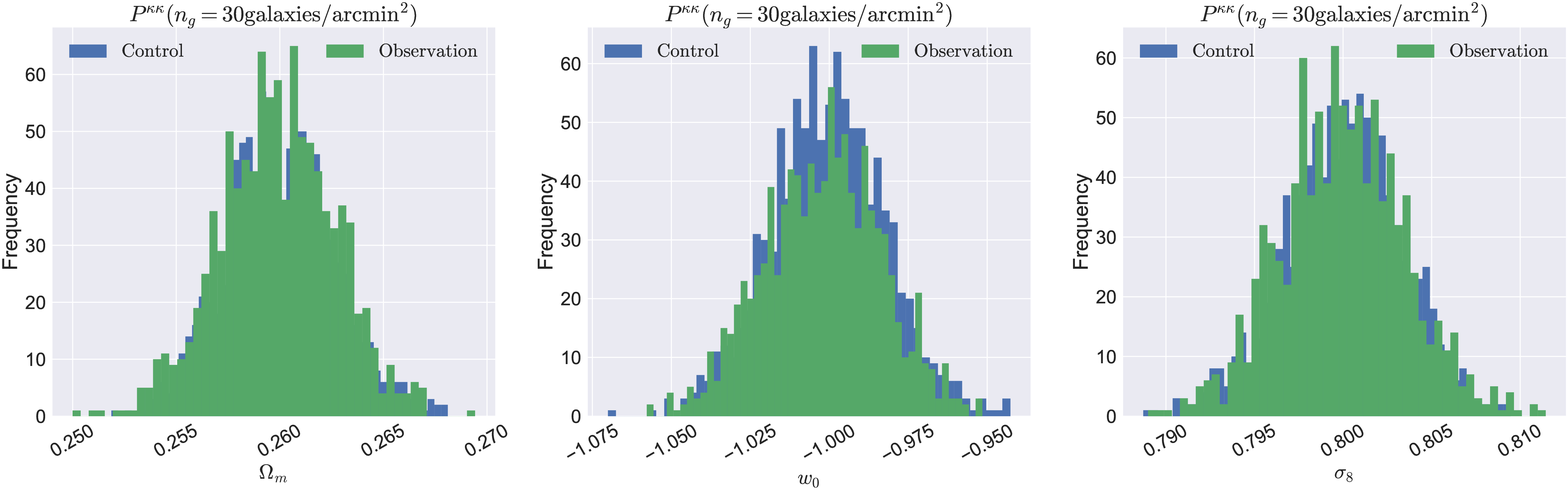}
\includegraphics[scale=0.3]{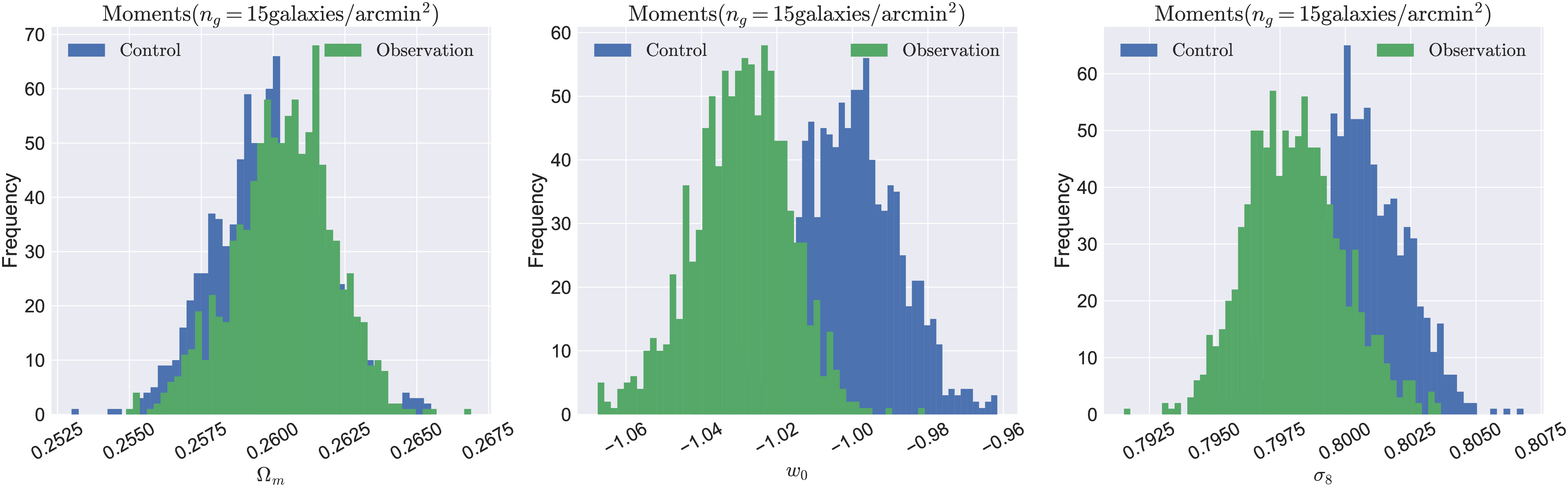}
\includegraphics[scale=0.3]{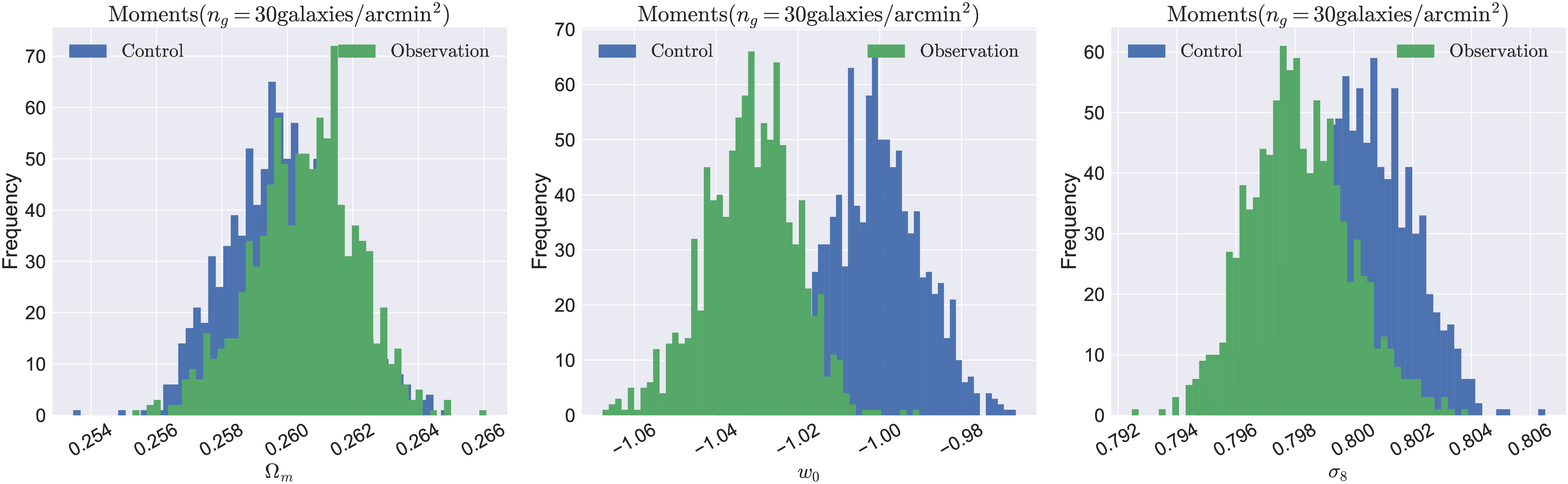}
\end{center}
\caption{Distributions of the parameter estimator (eq.~\ref{par:fisherest}) over 1000 realizations of an LSST--like survey. We show parameter estimates obtained using the $\kappa$ power spectrum (top panels) with shape noise added ($n_g=30\,{\rm galaxies/arcmin}^2$) and the $\kappa$ moments defined in eq.~(\ref{par:moments}) with shape noise added with both a galaxy density of 15 (middle panels) and 30 (bottom panels) galaxies/arcmin$^2$. In each panel, we show both a control case, in which the mock observation has been constructed with the Born approximation (blue), and the ``real'' case, in which the observation has been constructed with full ray--tracing (green). The $\kappa$ power spectrum from which the constraints in the upper panel have been derived, has been measured in 100 uniformly spaced bands with $\ell \in [150,10000]$. A Gaussian smoothing window of size $\theta_G=0.5\,{\rm arcmin}$ has been applied to the $\kappa$ maps prior to measuring the moments.}
\label{fig:parbias}
\end{figure*}

\begin{figure*}
\begin{center}
\includegraphics[scale=0.45]{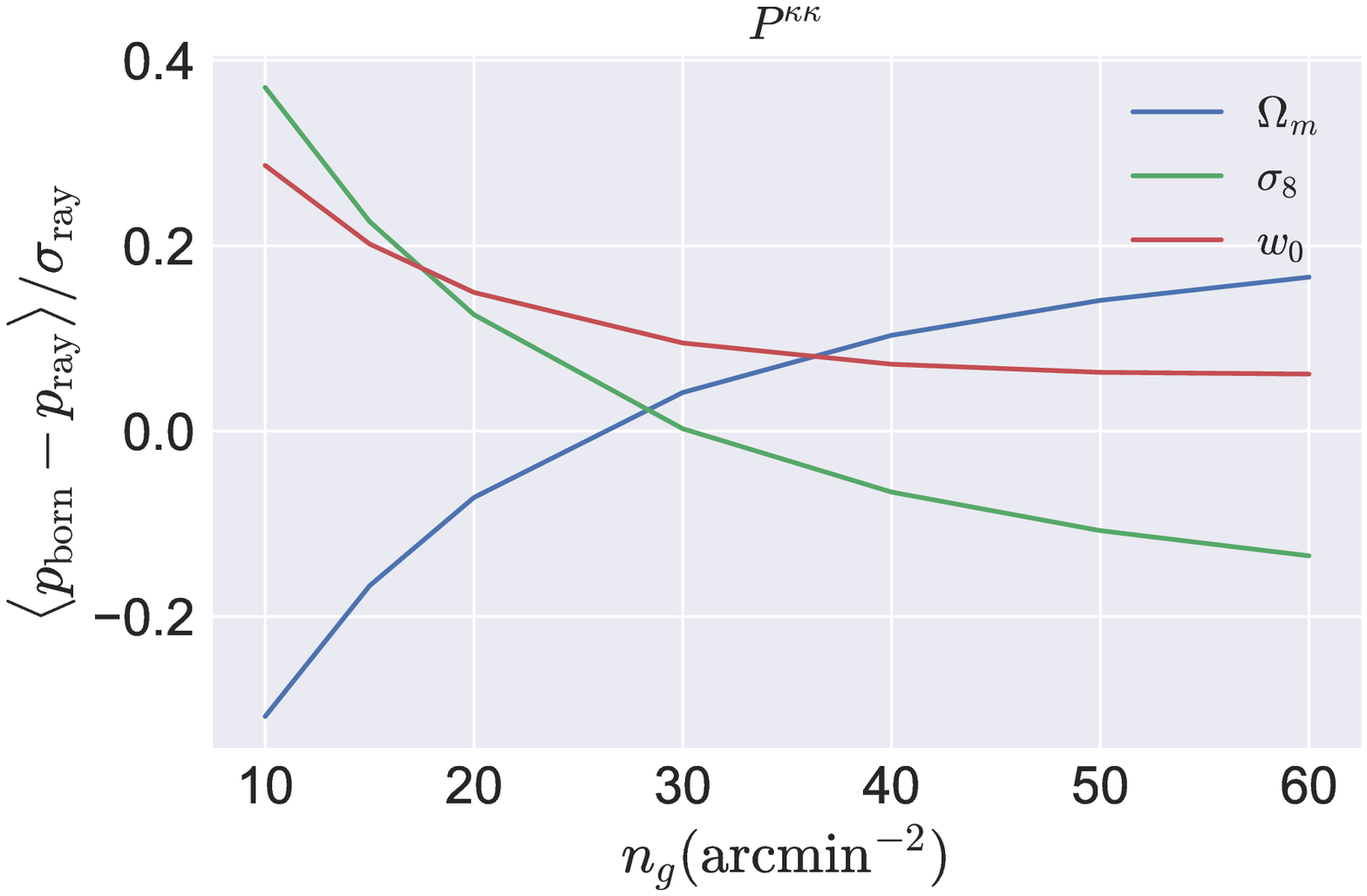} \includegraphics[scale=0.45]{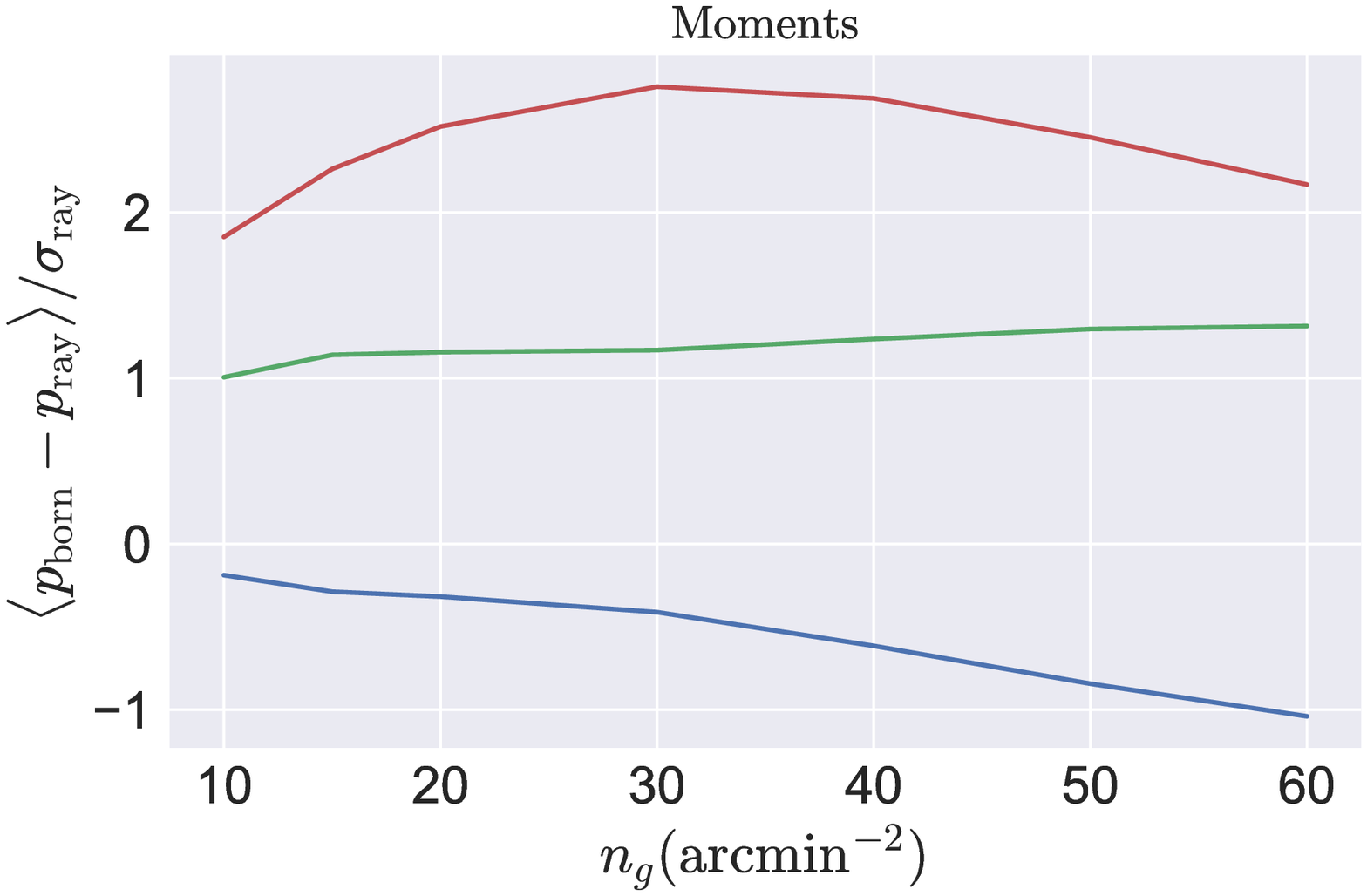}
\end{center}
\caption{Parameter bias, obtained from the $\kappa$ power spectrum (left panel) and the $\kappa$ moments (right panel), as a function of the galaxy density angular $n_g$ used to compute the shape noise rms in eq. (\ref{par:shaperms}). The plotted quantity is the difference between the mean of 1000 parameter fits using the estimator (\ref{par:fisherest}) with $\bb{M}$ measured with and without the Born approximation. The difference is shown in units of the estimator standard deviation. We show the trend for $\Omega_m$ (blue), $w_0$ (red) and $\sigma_8$ (green). The quoted $\sigma_{\rm ray}$ is calculated as a standard deviation. Parameter estimates PDFs are not Gaussian because the estimator $\hat{\mathbf{d}}_{\rm obs}$ which appears in equation (\ref{par:fisherest}) in general is non--Gaussian distributed. When we average over multiple fields of view though (this is the case for LSST), the Gaussianity assumption is justified by the central limit theorem. This is why the standard deviation $\sigma_{\rm ray}$ is a reasonable benchmark for statistical significance.}
\label{fig:parbiasSN}
\end{figure*}

\subsection{Forward model accuracy}

In this subsection, we compare the accuracy of the Born approximation in predicting WL image features, in particular we focus on the $\kappa$ power spectrum and the 9 moments described in equation (\ref{par:moments}). Figure \ref{fig:psResiduals} shows the auto power spectra of $\kappa_{\rm born},\kappa_{\rm ll},\kappa_{\rm geo}$ as defined in equation (\ref{sim:pb2}). At lowest order the lensing potential contributes to the $\kappa$ power spectrum quadratically. The first post--Born correction to $P^{\kappa\kappa}$ comes in at $O(\Phi^3)$ and is given by
\begin{equation}
\label{res:powercorr}
P^{\kappa\kappa} = P^{\kappa\kappa}_{\rm born,born} + 2P^{\kappa\kappa}_{\rm born,ll} + 2P^{\kappa\kappa}_{\rm born,geo} + O(\Phi^4),
\end{equation}
where we defined the $\kappa$ cross spectra as
\begin{equation}
\label{res:powercross}
\langle\tilde{\kappa}_a(\pmb{\ell})\tilde{\kappa}_b(\pmb{\ell}')\rangle = (2\pi)^2\delta^D(\pmb{\ell}+\pmb{\ell}')P_{a,b}^{\kappa\kappa}(\ell).
\end{equation} 
Figure \ref{fig:psResiduals} shows the magnitudes of the $O(\Phi^3)$ post--Born corrections to the convergence power spectrum. A similar analysis can be carried out for the various skewness and kurtosis moments of the $\kappa$ field. The main contributions to the skewness and kurtosis enter at $O(\Phi^3)$ and $O(\Phi^4)$, respectively. The first post--Born corrections, on the other hand, enter at $O(\Phi^5),O(\Phi^6)$. In more detail, we have

\begin{equation}
\label{res:skewcross}
\langle\kappa^3\rangle = \langle\kappa_{\rm born}^3\rangle + 3\langle\kappa_{\rm born}^2\kappa_{\rm ll}\rangle + 3\langle\kappa_{\rm born}^2\kappa_{\rm geo}\rangle + O(\Phi^5)
\end{equation}
and
\begin{equation}
\label{res:kurtcross}
\langle\kappa^4\rangle_c = \langle\kappa_{\rm born}^4\rangle_c + 4\langle\kappa_{\rm born}^3\kappa_{\rm ll}\rangle_c + 4\langle\kappa_{\rm born}^3\kappa_{\rm geo}\rangle_c + O(\Phi^6).
\end{equation}
Here the connected kurtosis components are defined as 
\begin{equation}
\label{res:kurtconnected}
\langle\kappa^3_a\kappa_b\rangle_c = \langle\kappa^3_a\kappa_b\rangle - 3\langle\kappa_a^2\rangle\langle\kappa_a\kappa_b\rangle.
\end{equation}

The magnitude of the post--Born corrections to the $\kappa$ skewness and kurtosis in equations (\ref{res:skewcross},\ref{res:kurtcross}) are shown in Figure \ref{fig:skResiduals}.
We also studied the accuracy of the Born approximation in predicting the 9 higher order convergence moments defined in eq.~(\ref{par:moments}). The results are shown in Figure \ref{fig:pdfMoments}.

\subsection{Parameter bias}
When using forward models based on the Born approximation to fit observations, parameter bias may occur if the forward model is not accurate enough. We studied if this is the case by simulating 1000 LSST--like observations in which the convergence field $\kappa(\pt)$ is constructed using the full ray--tracing procedure. We used the Born approximation to fit these mock observations using the Fisher parameter estimator in eq.~(\ref{par:fisherest}). The results are shown in Figure \ref{fig:parbias}. We also studied how our conclusions are influenced by the survey angular galaxy density $n_g$, which controls the amplitude of the shape noise in the $\kappa$ reconstruction. The scaling of the bias induced by the Born approximation on the power spectrum and $\kappa$ moments as a function of $n_g$ is shown in Figure \ref{fig:parbiasSN}.   


\section{Discussion}
\label{sec:discuss}
In this section, we discuss our main findings. Figure \ref{fig:csample} shows that the dominant post--Born correction to the $\kappa$ reconstruction comes from the geodesic term (eq.~\ref{sim:gp}). This can be seen by looking at the lens-lens term, which is over an order of magnitude smaller than the geodesic term, and by looking at the full $\kappa-\kappa_{\rm born}$ residuals, which show a very similar overall level, as well as structural detail, to the ones found in the $\kappa_{\rm geo}$ map. The reader might have noticed the curious alignment feature in Figure \ref{fig:csample}, in which all the dipolar structures in $\kappa_{\rm geo}$ appear to be horizontally aligned throughout the field of view. We believe that this alignment is not a numerical effect, but originates physically by the $\nabla\delta$ terms in eq.~(\ref{sim:gp}). Early in the line--of--sight integration, the field of view covers a very small section of the simulation box ($\sim7\,{\rm Mpc}$ for the first lens), within which dipolar structures are aligned. These alignments survive after the full integration is completed due to the $\nabla\delta\cdot\nabla\Phi$ term, which couples lenses at different redshifts. To provide evidence that the closest lenses to the observer are responsible for the alignment features, we performed the line--of--sight integration for $\kappa_{\rm geo}$ by rotating the first 5 lenses by 90 degrees. This is sufficient to coherently flip the dipolar structures from horizontal to vertical throughout the field of view.

Figure \ref{fig:psResiduals} shows a comparison between the Born--approximated $\kappa$ power spectra and the ones obtained with full ray--tracing, separating different orders in $\Phi$. We can conclude that the dominant contributions to the residuals come from the Born--geodesic cross power, which dominates the Born--lens--lens cross power by two orders of magnitude on small scales. The dashed blue line in Figure \ref{fig:psResiduals} corresponds to the residuals $P_{\rm ray} - P_{\rm born} - 2P_{\rm born,ll} - 2P_{\rm born,geo}$, which are comparable to the $O(\Phi^3)$ terms themselves. This means that the $O(\Phi^4)$ post--Born terms are completely overshot by cosmic variance and numerical noise in this case. Figure \ref{fig:skResiduals} shows that the geodesic and lens--lens contributions to the $\kappa$ skewness and kurtosis are comparable and that the lowest non--trivial post--Born corrections can fully account for the discrepancy between the Born--approximated quantities and the fully ray--traced ones. The geodesic and lens--lens corrections to the skewness we find agree, as orders of magnitude, with the ones already published in \citep{WLBispectrumDodelson}, although our measured reduced shear correction is larger. We stress, however, that our estimates and the ones in \citep{WLBispectrumDodelson} have been produced under different assumptions. In particular, \citep{WLBispectrumDodelson} use a fully analytical approach based on the Limber approximation, they consider different smoothing window functions and use their own assumptions for the bi--spectrum of $\kappa$. Figs. \ref{fig:psResiduals} and \ref{fig:skResiduals} indicate that truncating $\kappa$ at $O(\Phi^2)$ post--Born terms, leads to inaccurate predictions of the power spectrum and moments. In particular, higher-order terms involving $\langle \kappa_{\rm born}^n \kappa_{\rm ll}^m \kappa_{\rm geo}^2\rangle$ and higher, cause large deviations (as seen by the difference between the solid and dashed green curves) that, in an exact approach, are canceled by $O(\Phi^3)$ corrections to $\kappa$, as pointed out by \citep{HirataKrause,CMBPrattenLewis,ScoccimarroLoop}. For example, regarding the skewness of $\kappa$, the residuals $\kappa^3_{\rm born+geo}-\kappa^3_{\rm born}$ contain terms up to order $O(\Phi^6)$. Some of these high order terms should be canceled by $O(\Phi^3)$ corrections to the $\kappa$ image. These cancellations do not show up if one truncates the expansion of $\kappa$ to $O(\Phi^2)$. This is the reason behind the discrepancy between the green curves in Figure \ref{fig:skResiduals}. Therefore, in practice, in order to go beyond Born approximation, it is more advantageous, both for accuracy and for CPU time, to perform full ray--tracing (whose computational cost is comparable to those of the $O(\Phi^2)$ line--of--sight integrals; Table \ref{tab:benchmarks}).

In real observations, the convergence $\kappa$ is reconstructed via the Kaiser--Squires inversion procedure \citep{KaiserSquires}, applied on the reduced shear $\pmb{\gamma}/(1-\kappa)$. In order to avoid possible biases in parameter inferences, one needs to take this into account. This can be done within the Born approximation, as the quantity $\kappa\pmb{\gamma}$ is the dominant $O(\Phi^2)$ reduced shear correction, which can be calculated from the Born $\kappa$ images and is independent from the line of sight integration procedure. The $O(\Phi^2)$ reduced shear correction to $\kappa$ is easily incorporated in the forward models.

Figure \ref{fig:pdfMoments} shows the PDF of the 9 $\kappa$ moments defined in eq.~(\ref{par:moments}) over 1000 realizations of an LSST--like survey. This figure clearly shows that, while the Born approximation is a good predictor of the quadratic moments $\pmb{\mu}_2$ within one standard deviation, the higher moments $\pmb{\mu}_3,\pmb{\mu}_4$ show 2 to 3$\sigma$ deviations from their Born--approximated counterparts. This can lead to parameter bias when fitting observations. Figure \ref{fig:parbias} studies this possibility and shows that the Born approximation is sufficient to fit cosmology with the power spectrum of an LSST--like survey, with the result holding also for galaxy densities as high as $60\,{\rm galaxies/arcmin}^2$, as can be seen in Figure \ref{fig:parbiasSN}. The accuracy requirements, however, might be stricter for deeper surveys, such as Euclid \citep{Euclid}, which requires systematic effects not to be greater than $\sigma/3$. While the Born approximation is accurate in predicting the $\kappa$ power spectrum, the same is not true for higher order $\kappa$ moments. We have previously shown that these moments contain significantly more information than the power spectrum \citep{MinkPetri}, an expectation that was confirmed when fitting CFHTLenS data~\citep{CFHTMink}. The middle and bottom row of panels in Figure \ref{fig:parbias} show that indeed the constraints from the moments are $\approx$ 2 times tighter than from the power spectrum (seen by the narrow widths of the PDFs). As a result, highly significant biases are observed when fitting an observation with Born--approximated $\kappa$ moments. Although the presence of Gaussian galaxy shape noise reduces the significance of this bias, it cannot completely eliminate it, as higher order moments are sensitive to non--Gaussian statistical information in $\kappa$ images. 
These results lead us to the conclusion that the Born approximation is sufficient for future WL analyses that use the $\kappa$ power spectrum to constrain cosmology. This approach also has the advantage of being 4 times computationally faster than full ray--tracing, as we show in Table \ref{tab:benchmarks} for a selected reference test case. 

A few technical considerations are in order here. When adopting the Born approximation, it seems tempting to collapse all the particles in the $N$--body outputs in a single lens plane transverse to the line of sight, assigning to each particle a weight $W(\chi,\chi_s)$. Unfortunately the function $W$ is concave in $\chi$ and this ``single--lens--plane'' approach under--estimates $\kappa$ by a non--negligible amount. In order to avoid this concavity effect we need to use multiple discrete steps. We do not need to perform the full ray--tracing calculations however, but we can simply add the density values on the lens planes as light rays travel between them. The Born algorithm scales as $O(N_l)$. Ray--tracing has the same big $O$ complexity, but takes more time because at each step in the integration one needs to compute $O(N_R)$ second derivatives and $2\times 2$ matrix products, where $N_R$ is the number of light rays that resolve the $\kappa$ image. The quadratic corrections to $\kappa$ can also be computed in $O(N_l)$ when appropriate caching is used, but has a slightly different runtime compared to ray--tracing due to the different structure of the linear algebra operations involved. The memory usage is regulated by the number of two dimensional grids that need to be cached in order to perform the integration steps: in the Born case just a density grid is needed, but in the other cases one needs to keep track of the intermediate $\Phi$ derivatives as well.

When forward modeling higher order moments of $\kappa$ full ray--tracing is required in order to obtain unbiased constraints.  


\section{Conclusions}
\label{sec:conclude}
In this work, we used cosmological simulations to study the effectiveness of the Born approximation in predicting WL observables, and to compute the corresponding biases on parameter constraints. Our main findings can be summarized as follows:

\begin{itemize}
\item The post--Born corrections to the convergence power spectrum, skewness and kurtosis are well explained by the next--to--leading orders in the $\Phi$ expansion of each statistic. 
\item Using the \ttt{LensTools} software package, Born integration costs 4 times less than ray--tracing, and consumes about half as much memory.
\item The Born approximation for the $\kappa$ power spectrum leads to negligible parameter bias for an LSST--like survey, and holds for survey galaxy densities as high as $60\,{\rm galaxies/arcmin^2}$.
\item Fitting an observation with Born--approximated higher $\kappa$ moments leads to significant bias in the $(\Omega_m,w_0,\sigma_8)$ triplet, even in the presence of galaxy shape noise.    
\end{itemize}
In this work we examined the validity of the Born approximation for WL galaxy surveys, but a similar study could in principle be carried for the lensing potential reconstruction from CMB temperature and polarization data \citep{CMBCalabrese}. Lensing of the CMB probes structures over a wider range of redshifts and hence CMB lensing observables can be expected to closer to Gaussian than their galaxy lensing counterparts. This suggests the possibility that non--Gaussian features in CMB lensing data could come from post--Born corrections of $O(\Phi^2)$ rather than from intrinsic non--Gaussianity in the Born $O(\Phi)$ terms. This possibility has been suggested by \citep{CMBPrattenLewis}, who looked at the CMB lensing $\kappa$ bi--spectrum. Because of the high significance with which non--Gaussianity in the CMB lensing potential can be detected with future Stage IV experiments \citep{CMBLNG}, we propose to investigate post--Born corrections to CMB lensing observables in future work.  


\section*{Acknowledgments}
We thank the anonymous referee for the useful and insightful comments. We thank Uro\v{s} Seljak and Colin Hill for useful discussions. The $N$--body simulations in this work were performed at National Energy Research Scientific Computing Center (NERSC). We thank the LSST Dark Energy Science Collaboration (DESC) for the allocation of time. The ray--tracing simulations and parameter estimation calculations were performed at the NSF XSEDE facility, supported by grant number ACI-1053575, and at the New York Center for Computational Sciences, a cooperative effort between Brookhaven National Laboratory and Stony Brook University, supported in part by the State of New York. This work was supported in part by the U.S. Department of Energy under Contract No. DE-SC00112704, the NSF Grant No. AST-1210877 (to Z.H.), the Research Opportunities and Approaches to Data Science (ROADS) program at the Institute for Data Sciences and Engineering at Columbia University (to Z.H.) and a Simons Fellowship in Theoretical Physics (to Z.H.).


\bibliography{ref}

\label{lastpage}
\end{document}